\documentclass[prb,showpacs,twocolumn,superscriptaddress,floatfix]{revtex4}
\usepackage[english]{babel}
\usepackage{graphicx}
\usepackage{amsmath}
\usepackage{amssymb}
\begin{document}
\title{Effect of spin-orbit coupling on the excitation spectrum of Andreev billiards}
\author{B. B{\'e}ri}
\affiliation{Department of Physics of Complex Systems,
E{\"o}tv{\"o}s University,
 H-1117 Budapest, P\'azm\'any P{\'e}ter s{\'e}t\'any 1/A, Hungary}
\author{J.~H.~Bardarson}
\affiliation{Instituut-Lorentz, Universiteit Leiden, P.O. Box 9506,
2300 RA Leiden, The Netherlands}
\author{C.~W.~J.~Beenakker}
\affiliation{Instituut-Lorentz, Universiteit Leiden, P.O. Box 9506,
2300 RA Leiden, The Netherlands}
\date{December, 2006}
\begin{abstract}
We consider the effect of spin-orbit coupling on the low energy excitation
spectrum of an Andreev billiard (a quantum dot weakly coupled to a
superconductor), using a dynamical numerical model (the spin Andreev
map). Three effects of spin-orbit coupling are obtained in our
simulations:
In zero magnetic field: (1) the narrowing of the distribution of the
excitation gap; (2) the appearance of oscillations in the average density of states.
In strong magnetic field: (3) the appearance of a peak in the average density of states at
zero energy. All three effects have been predicted by random-matrix
theory.
\end{abstract}
\maketitle{}

\section{Introduction}
\begin{figure}[tb]
  \begin{center}
    \includegraphics[width=0.6\columnwidth]{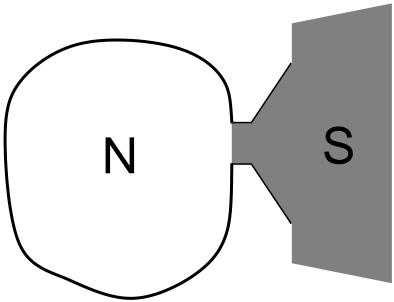} 
  \end{center}
  \caption{Sketch of an Andreev Billiard: A quantum dot (N) connected to a
  superconductor (S) by a point contact. Spin-orbit coupling is present in the quantum
  dot.}
  \label{fig:setup}
\end{figure}

A  quantum dot in a two-dimensional electron gas has a mean level spacing which is independent of energy and depends only on geometrical factors (area) and material properties (effective
mass). The nature of the electron dynamics (chaotic versus
integrable) and the presence or absence of symmetries, such as time-reversal
and spin-rotation symmetry, have no effect on the mean density of states. The
situation changes if the quantum dot is 
coupled to a superconductor (see
Fig.~\ref{fig:setup}). The presence of the superconductor strongly affects
the excitation spectrum of such an Andreev billiard. The density of states  at
the Fermi level is suppressed in a way which is sensitive to the nature of the
dynamics and existing symmetries.\cite{Bee05} 
While the effect of broken time-reversal symmetry on the density of states has
been studied extensively,\cite{Alt96,Fra96,Mel97,Goo05} the effect of broken spin-rotation
symmetry due to spin-orbit coupling has only been partially investigated.\cite{Alt97,Cht03,Dim06}

In this paper we study in computer simulations three effects of the spin-orbit
coupling. All of these effects have been predicted by random-matrix theory
(RMT),\cite{Tra96,Alt97,Vav01} but have so far not been confirmed in a dynamical
model. The first two of
these effects,\cite{Tra96,Vav01} present in the absence of a magnetic field, are the reduction of the sample-to-sample fluctuations of the
excitation gap and the appearance of oscillations as a function of energy in the average
density of states. The third effect 
appears in a magnetic field strong enough to close the excitation gap.  While
in the absence of  spin-orbit coupling the average density of states vanishes at the
Fermi level, in the presence of spin-orbit coupling 
it peaks at the Fermi level at twice the value in the
normal state.\cite{Alt97} 

\section{Predictions of random-matrix theory}
\label{sec:RMT}
We begin by briefly summarizing the RMT of the Andreev billiard.\cite{Bee05} 
In perturbation theory the density of states of non-degenerate levels (in zero or weak magnetic field) has a square root dependence on
energy near the gap,\cite{Mel96,Mel97}
\begin{equation}
  \label{eq:rhoMF}
  \rho_\text{pert}(E) = \frac{1}{\pi}\sqrt{\frac{E-E_\text{gap}}{\Delta_\text{gap}^3}},\quad E\rightarrow E_\text{gap}.
\end{equation}
The parameters $E_\text{gap}$ and $\Delta_\text{gap}$ are given by
\begin{equation}
  \label{eq:Egap}
  E_\text{gap}=c E_T,\quad \Delta_\text{gap}
  = \left(\frac{s\delta}{2}\right)^{2/3}\left(\frac{dE_T}{4\pi^2}\right)^{1/3}\!\!\!.
\end{equation}
Here $\delta = 2\pi\hbar^2/mA$ is the mean level spacing in the isolated quantum dot (area $A$, effective mass $m$), $N =
{\rm Int}[k_FW/\pi]$ is the
number of modes in the ballistic point contact (width $W$, Fermi wave vector $k_F$) connecting it to the superconductor, and $E_T =
N\delta/4\pi$ is the Thouless energy. These parameters refer to two-fold degenerate levels and modes, corresponding to $s=2$. If both
spin-rotation and time-reversal symmetries are broken, the two-fold degeneracy is lifted and one should take $s = 1$. The numerical
coefficients $c$ and $d$ are magnetic field dependent. For $B=0$ one has
\begin{equation}
   \label{eq:cdparams}
  c=2\gamma^{5/2}, \quad d=(5-2\sqrt{5})\gamma^{5/2},
\end{equation}
with $\gamma =(\sqrt{5}-1)/2$ the reciprocal of the golden ratio. For $B \neq 0$ they should be calculated from the RMT solution given
in Ref.~\onlinecite{Mel97}.
  
The perturbation theory has $\delta/E_T$ as a
small parameter and gives the density of states with an energy
resolution of order $E_T$, which is
a macroscopic energy scale. Since spin-orbit effects typically appear as quantum
corrections, no sign of spin-orbit coupling can be seen in
such a calculation. 
In particular, the magnetic field dependence of
the perturbative density of states is the same with and without spin-orbit
coupling --- the only difference being that in a magnetic field, in the absence of
spin-rotation symmetry, there is no level degeneracy thus $\Delta_{\text{gap}}$ is
$2^{2/3}$ times smaller than in the spin-rotation symmetric case.

\begin{figure}[tb]
  \begin{center}
    \includegraphics[width=0.95\columnwidth]{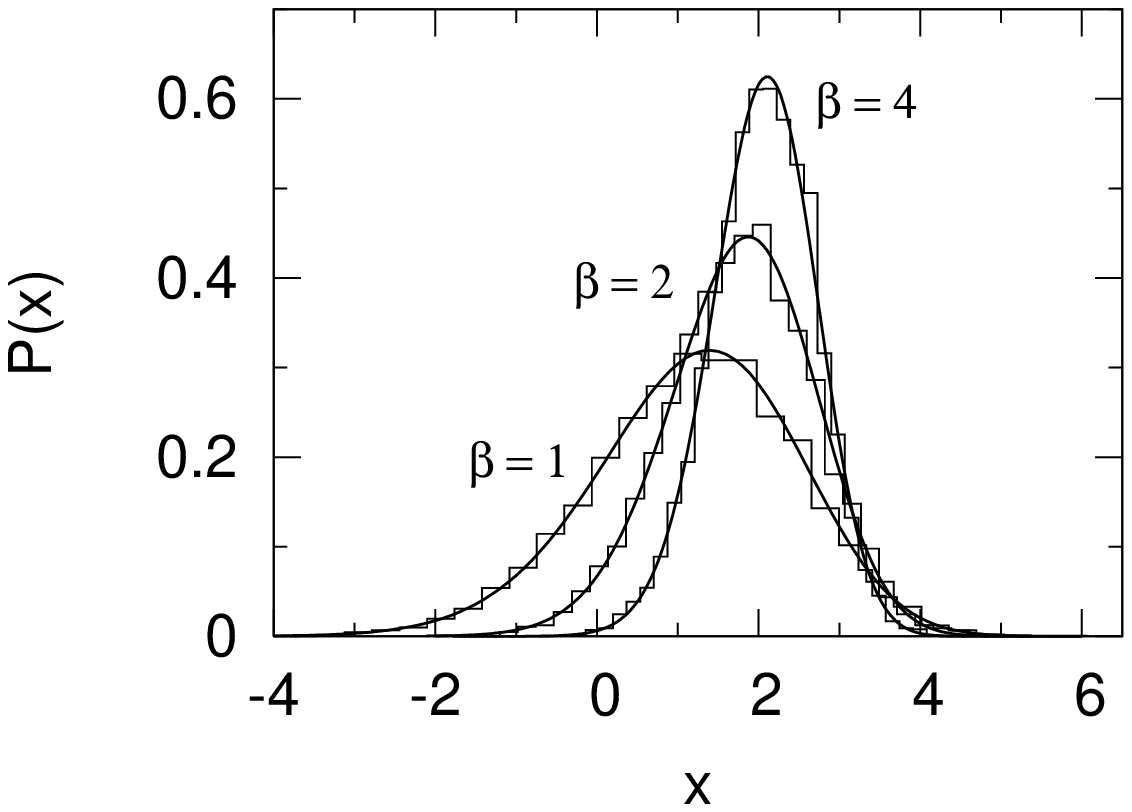} 
    \includegraphics[width=0.95\columnwidth]{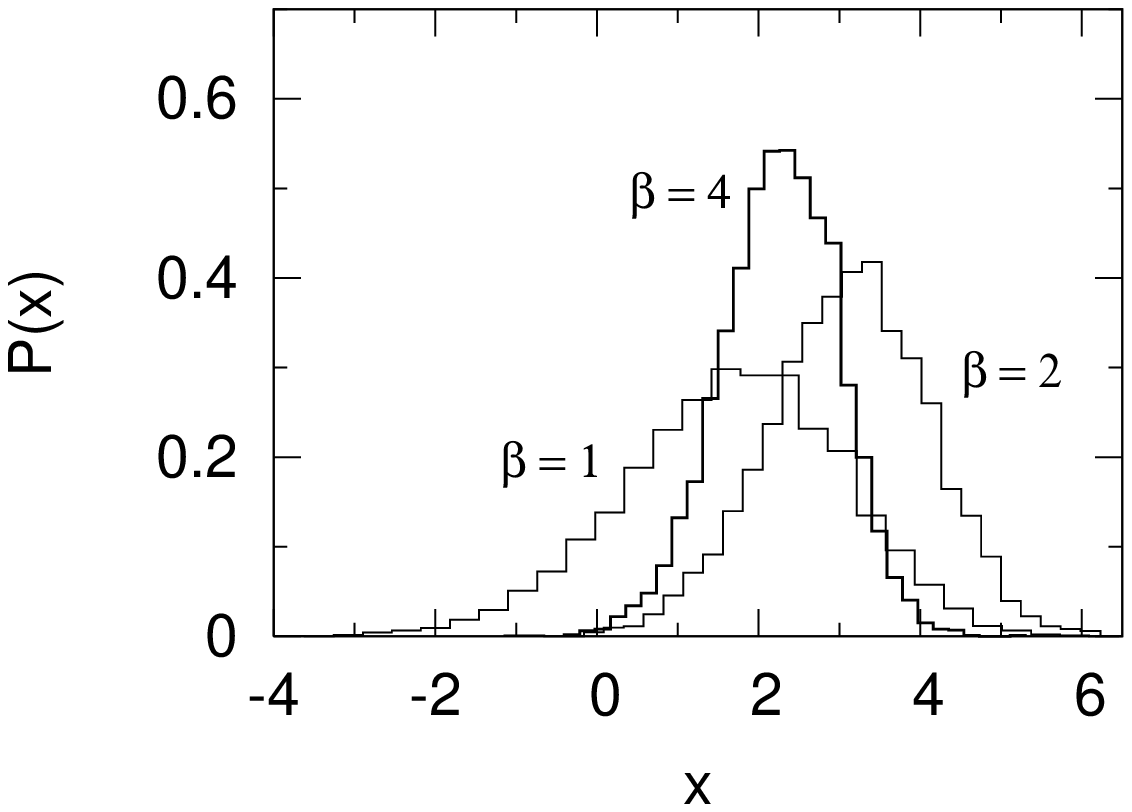}
   \end{center}
   \caption{Probability distribution of the rescaled excitation gap $x=(E_{1}-E_\text{gap})/\Delta_\text{gap}$. Smooth curves (upper
   panel) are the predictions of random-matrix theory, for the three symmetry classes $\beta=1,2,4$. Histograms are the results of the
   numerical simulation of the spin Andreev map, without spin-orbit coupling (zero magnetic field, $\beta=1$) and with spin-orbit
   coupling (zero magnetic field, $\beta=4$; weak magnetic field, $\beta=2$). The histograms in the lower panel are plotted without any
   fitting, while in the upper panel $E_\text{gap}$ and $\Delta_\text{gap}$ are treated as fit parameters.
}
  \label{fig:P(x)}
\end{figure}

To capture the spectral properties on the mesoscopic energy scale of
order $\Delta_\text{gap}$, one needs to go beyond perturbation theory. According to
the universality hypothesis of Vavilov {\it et al.\ }\cite{Vav01} the probability
distribution of the lowest level $E_1$ in properly scaled units is
universal and identical to the distribution of the smallest eigenvalue of
random Gaussian matrices from the three symmetry classes of RMT.
The appropriate scaling is in terms of the dimensionless variable
$x_1=(E_1-E_\text{gap})/\Delta_\text{gap}$ and the universal distributions are given by\cite{Tra96,Ede05}
\begin{equation}
\label{eq:probdist}  
 P_\beta(x_1)=-\frac{d}{dx_1}F_\beta(x_1),
\end{equation}
where
\begin{subequations}
\begin{align}
F_1(x)&=\sqrt{F_2(x)}\exp \left(-\frac{1}{2}\int_{-\infty}^x q(x')\,dx'\right),\\
F_2(x)&=\exp \left( -\int_{-\infty}^x (x-x')q(x')^2\,dx' \right), \\
F_4\left(\frac{x}{2^{2/3}}\right) &= \sqrt{F_2(x)}\cosh
\left(\frac{1}{2}\int_{-\infty}^x q(x')\,dx'\right).
\end{align}
\end{subequations}
The function $q(x)$ is the solution of the differential equation
\[q''(x)=-xq(x)+2q^3(x). \]
The boundary condition is $q(x)\rightarrow {\rm Ai}(-x)$ as \mbox{$x\rightarrow
-\infty$}, with ${\rm Ai}(x)$ being the Airy function. 
The three distributions
are plotted in Fig.~\ref{fig:P(x)} (top panel).
The symmetry index $\beta$ takes values $\beta=1$ for time-reversal and
spin-rotation invariant systems, $\beta=4$  when time-reversal symmetry is
present but spin-rotation symmetry is broken, and $\beta=2$ for systems with
broken time-reversal symmetry.

Near the gap, the average density of states in terms of the variable
$x=(E-E_\text{gap})/\Delta_\text{gap}$ is given by\cite{FNH,Tra05}
\begin{subequations}
\begin{align} 
\rho_1(x)=\rho_2(x)&+ \tfrac{1}{2}{\rm Ai}(-x)\left[1-\int_{-x}^{\infty} {\rm Ai}(y)dy \right],\\
\rho_2(x)&=x{\rm Ai}^2(-x)+[{\rm Ai}'(-x)]^2,\\
2^{1/3}\rho_4\left(\frac{x}{2^{2/3}}\right)&=\rho_2(x)-\frac{1}{2}{\rm Ai}(-x)\int_{-x}^{\infty} {\rm Ai}(y)dy.
\end{align}
\end{subequations}

The distribution $P_2$ and the density of states $\rho_2$ are applicable in an intermediate magnetic field range,
which exists because the flux needed to close the gap is much larger than the
flux needed to break the time-reversal symmetry. For intermediate fluxes (\mbox{$\Phi \gtrsim 
(h/e)N^{1/3}\sqrt{\tau_\text{erg}\delta/\hbar}$} with
$\tau_\text{erg}=\sqrt{A}/v_{\rm F}$ the ergodic time and $v_{\rm F}$ the Fermi
velocity), there will still be a gap, but its
fluctuations  are governed by the $\beta = 2$ symmetry class. In
this case the presence or absence of spin-orbit coupling only affects the parameter $\Delta_\text{gap}$ (which is reduced by a factor $2^{2/3}$ in the
absence of spin-orbit coupling, because the level degeneracy parameter goes from $s=2$ to $s=1$); the gap distribution $P_2$ and the density of states $\rho_2$ in rescaled variables do not depend on the
presence or absence of spin-rotation symmetry.

If the flux is made much larger ($\Phi \gg (h/e) \sqrt{N\tau_\text{erg}\delta/\hbar}$),
such that the gap closes, spin-orbit coupling starts to play a role again. The reason is that an energy level $E$ and its mirror level at
$-E$ can repel each other, and this repulsion depends on the presence or absence of spin-orbit coupling. When there is still a
gap these levels are widely separated and this repulsion is not effective.\cite{Alt97}

According to Altland and Zirnbauer,\cite{Alt97} the RMT of an Andreev billiard in strong magnetic field is in a new symmetry class
called C (D) in the absence (presence) of
spin-orbit coupling. The average density of states in these symmetry classes is
\begin{equation}
  \label{eq:rhoCD}
  \rho_{\pm}(E) = \frac{4}{s\delta}\left[1\pm\frac{\sin(8\pi E/s\delta)}{8\pi E/s\delta}\right].
\end{equation}
The minus (plus) sign should be taken for symmetry class C (D). This result
 expresses the fact that in the absence  of spin-orbit
coupling, a level and its mirror level repel each other leading to a vanishing
 density  of states at $E=0$. In
the presence of spin-orbit coupling the repulsion disappears and
 levels pile up at the Fermi level, leading to a peak in the  density of
states at $E=0$.

\section{Spin Andreev map}
To verify these predictions of RMT in a dynamical model we combine the general construction of an Andreev map\cite{Jac03} with the spin kicked
rotator.\cite{Sch89,Bar05}
The starting point of our discussion is the spin generalized Bogoliubov-De Gennes Hamiltonian
\cite{DeGennes}
\begin{equation}
  \label{eq:BdG}
  \mathcal{H}_{\text{BdG}} = 
   \begin{pmatrix}
     H-E_{\rm F} & \Delta \\
     \Delta^* & E_{\rm F}-\mathcal{T}H\mathcal{T}^{-1}
   \end{pmatrix}.
\end{equation}
Here $H$ is the single particle Hamiltonian, $E_{\rm F}$ is the Fermi energy,
and $\Delta$ is the superconducting pair-potential. The operator $\mathcal{T}$
stands for time-reversal, and will be specified later.   

With the Bogoliubov-De Gennes Hamiltonian as a guide we construct the 
spin Andreev map. First we note that if an electron in the normal metal evolves with time-evolution operator $F(t)$, the hole evolves
with the transformed time-evolution operator $\mathcal{T}F(t)\mathcal{T}^{-1}$. Second, since we are interested in low energy
phenomena, only the dynamics on long time scales is important. On time scales
much larger than $\tau_{\rm erg}$, the
dynamics can be described as a mapping on a two-dimensional Poincar\'e surface
of section. This amounts to a stroboscopic description
where we are only concerned with the state of the electron when it bounces off
the boundary. 

The quantum map we use is the computationally efficient
spin kicked rotator, given in terms of a Floquet matrix,\cite{Sch89, Bar05} 
\begin{equation}
  \label{eq:Floquet}
F_{ll'}=e^{i\varepsilon_0}(\Pi U X U^\dagger \Pi)_{ll'}, \quad l,l' = 0,1,\ldots, M-1.
\end{equation}
The integer $M$ sets the level spacing $\delta = 2\pi/M$.
The $M \times M$ matrices appearing in Eq.~\eqref{eq:Floquet} have quaternion matrix
elements, and are given  by
\begin{subequations}
\begin{align}
\Pi_{ll'}&=\delta_{ll'}e^{-i\pi (l+l_0)^2/M }\sigma_0, \\ 
U_{ll'}&=M^{-1/2}e^{-i2\pi ll'/M}\sigma_0,\\
X_{ll'}&= \delta_{ll'}e^{-i(M/4\pi)V(2\pi l/M)},
\end{align}
\end{subequations}
with 
\begin{equation}
V(\theta) =  K\cos(\theta+\theta_0)\,\sigma_0 + K_\text{so}(\sigma_1\sin2\theta + \sigma_3\sin\theta).
\end{equation}
The quaternions are represented using the Pauli matrices $\sigma_i$ with $\sigma_0$ the $2 \times 2$ unit matrix. The matrix $X$ corresponds to the
spin-orbit coupled free motion inside the dot 
and $\Pi$ gives scattering off the boundaries of the dot. This map is
classically  chaotic for kicking strength $K \gtrsim 7.5$. The parameter 
$K_\text{so}$ breaks spin-rotation symmetry, $\theta_0$ breaks time-reversal symmetry and $l_0$ breaks other symmetries of the map. The spin-orbit coupling time $\tau_\text{so}$
(in units of the stroboscopic period $\tau_0 \approx \tau_{\rm erg}$) is related to
$K_\text{so}$ through  
$\tau_\text{so} = 32\pi^2/(K_\text{so}M)^2$. 
The parameter $\varepsilon_0$ corresponds to the Fermi energy.
In the above representation of the Floquet matrix, 
the time-reversal operator is given by ${\cal T}=i \sigma_2 {\cal K}$
where ${\cal K}$ is the operator of complex conjugation.\cite{Bar05} Therefore, the hole Floquet matrix
is given  by $\sigma_2 F^* \sigma_2 \equiv \bar{F}$,  where the overbar denotes
quaternionic complex conjugation.

The spin Andreev map is constructed from the electron and hole Floquet
matrices in the same way as in the absence of spin-orbit coupling,\cite{Jac03} 
\begin{subequations}
\begin{align}
  \label{eq:aKR}
  \mathcal{F} &= \mathcal{P}
  \begin{pmatrix}
    F & 0 \\ 0 & \bar{F}\\
  \end{pmatrix},\\
  \mathcal{P} &= 
  \begin{pmatrix}
    1-P^TP & -iP^TP \\
    -iP^TP & 1-P^TP
  \end{pmatrix}.
\end{align}
\end{subequations}
The projection matrix $P$ projects onto the contact with the superconductor. Its matrix elements are $P_{kl} =
\delta_{kl}\sigma_0\sum_{i=1}^N\delta_{l,n_i}$ where the set of indices $\{ n_i \}$
corresponds to the modes coupled to the superconductor.
The dwell time is therefore $\tau_{\rm dwell}=M/N$. The corresponding Thouless energy is $E_T = N\delta/4\pi = (2\tau_\text{dwell})^{-1}$. 
As shown in Ref.~\onlinecite{Bar05}, the magnetic field scale at which the gap closes is given by $\theta_c = 4\pi\sqrt{N}/(KM^{3/2})$.
From the definitions in Eq.~\eqref{eq:Egap} the scaling parameters in the spin Andreev map become
\begin{equation}
\label{eq:scalingpar}
  E_\text{gap} = \frac{cN}{2M}, \quad
  \Delta_\text{gap} = \frac{(s^2dN)^{1/3}}{2M}.
\end{equation}

The Floquet matrix has the symmetry
\begin{equation}
  \label{eq:Symmetry}
  \mathcal{F} = {\cal CT} \mathcal{F}({\cal CT})^{-1}, \quad 
  {\cal C} = \begin{pmatrix}
    0 & -1 \\
    1 & 0
  \end{pmatrix},
\end{equation}
corresponding to the ${\cal CT}$-antisymmetry of $\mathcal{H}_{\text{BdG}}$, the fundamental discrete symmetry of normal-superconducting systems.\cite{Alt97}
The eigenphases of the Floquet matrix $\mathcal{F}$, defined as the solutions
of 
\begin{equation} 
  \label{eq:eqseq}
  \det(\mathcal{F} - e^{-i\varepsilon})=0, 
\end{equation}
play the role of the discrete excitation energies in the Andreev billiard. From the
symmetry~\eqref{eq:Symmetry} it follows that they come in pairs, 
$\pm\varepsilon$,  as required. 

\section{Numerical results and comparison with RMT}

\begin{figure}[tb]
  \begin{center}
    \includegraphics[width=0.95\columnwidth]{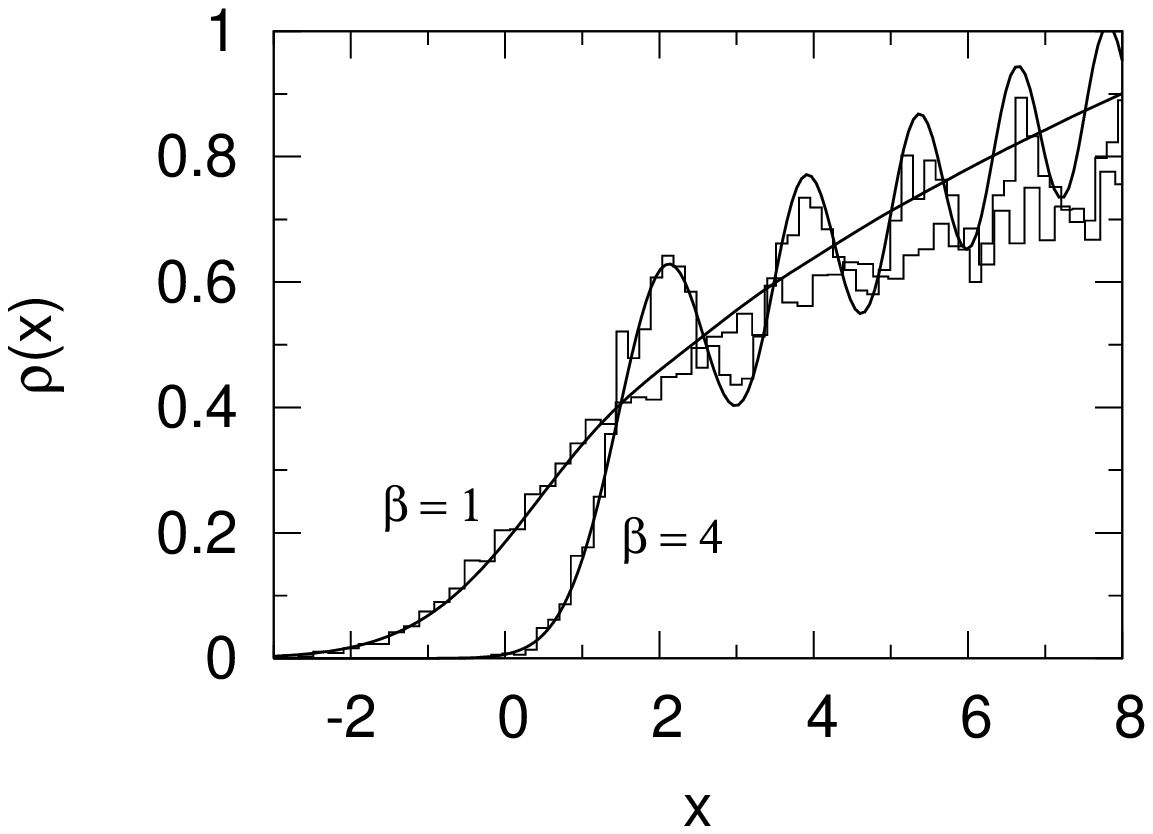} 
    \includegraphics[width=0.95\columnwidth]{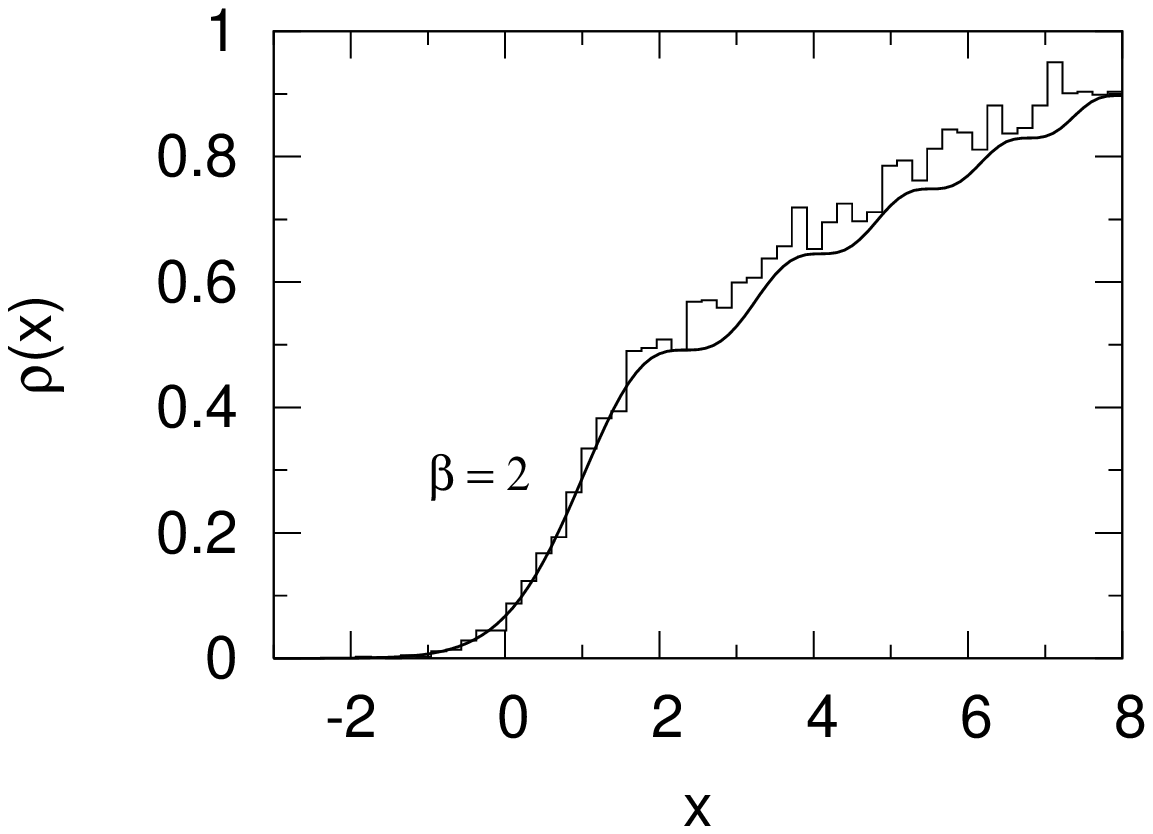}
   \end{center}
   \caption{Average density of states in rescaled energy units, $x=(E-E_\text{gap})/\Delta_\text{gap}$, in zero magnetic field (upper
   panel, $\beta=1,4$) and in a weak magnetic field (lower panel, $\beta=2$). The smooth curves are the RMT predictions, the histograms
   are the results of the simulation (using the same values of the fit parameters $E_\text{gap},\Delta_\text{gap}$ as in the upper panel
   of Fig.~\ref{fig:P(x)}).}
  \label{fig:rhoscale(x)}
\end{figure}

\begin{figure}[tb]
  \begin{center}
    \includegraphics[width=0.6\columnwidth,angle=270]{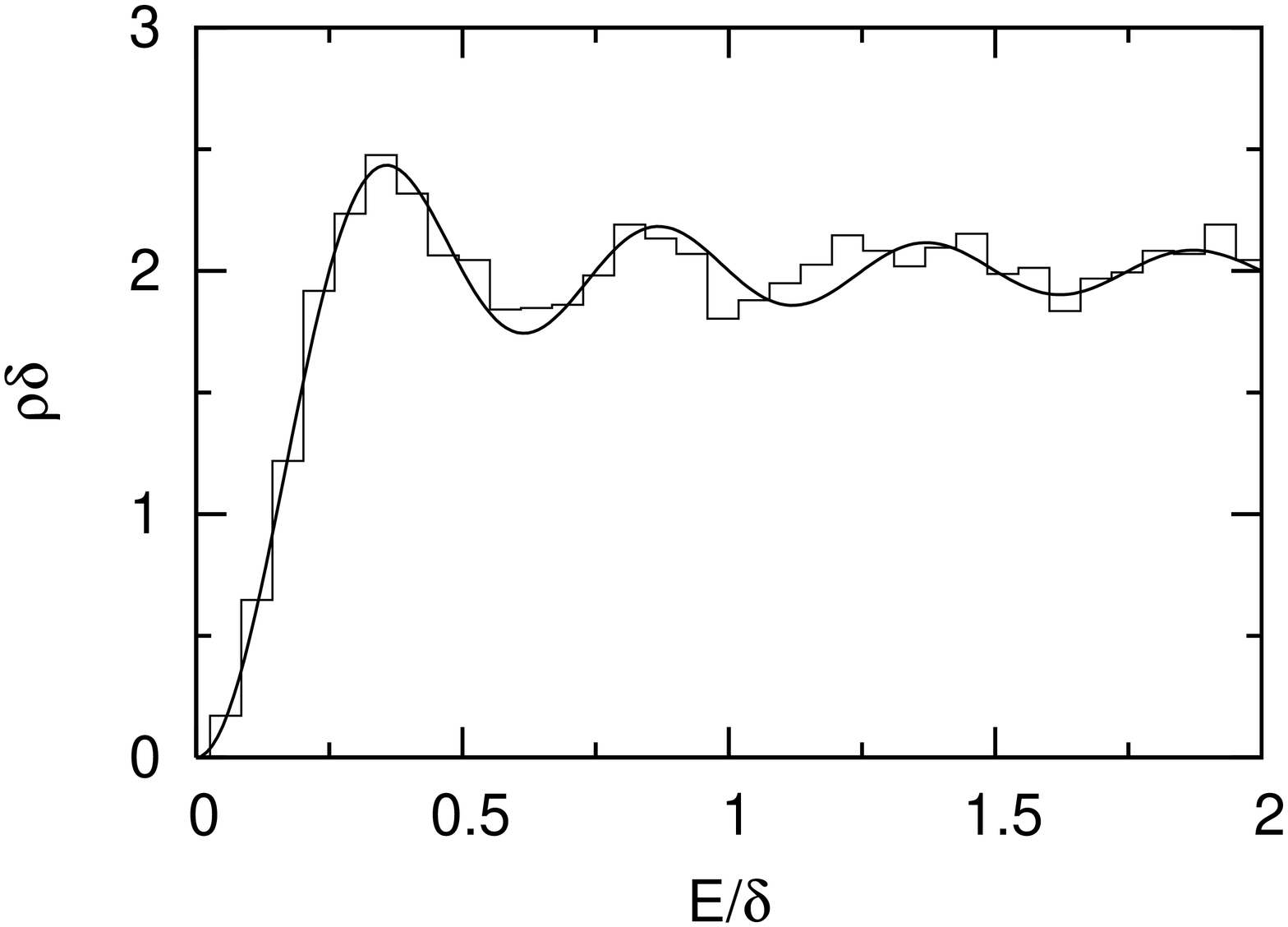} 
    \includegraphics[width=0.6\columnwidth,angle=270]{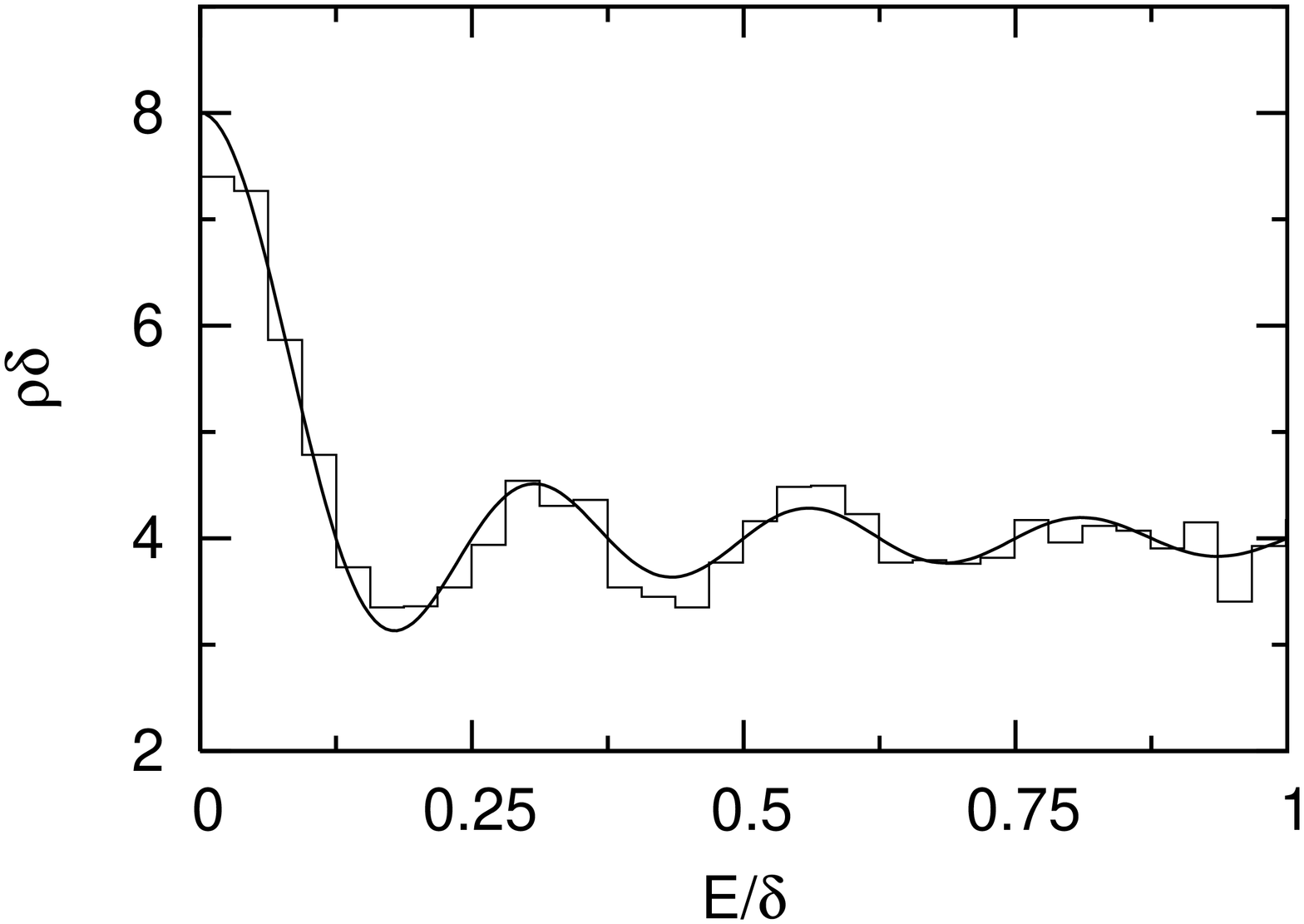}
  \end{center}
  \caption{Average density of states in a magnetic field sufficiently strong to close the excitation gap. Upper panel: without spin-orbit
  coupling; Lower panel: with spin-orbit coupling. The smooth curves are the RMT predictions from Eq.~\eqref{eq:rhoCD} and the histograms are the
  results of the simulation (now without any fit parameters).}
  \label{fig:rhoCD}
\end{figure}

In Fig.~\ref{fig:P(x)} we plot the excitation gap distribution (histograms) from our numerical simulation with parameters $K=41.123$,
$M=4096$, $N=205$. The smallest $\varepsilon$ solving Eq.~\eqref{eq:eqseq} was calculated for some 6000 different Fermi energies and
positions of the contact to the superconductor. To generate the three symmetry classes we took: $\theta_{0}/\theta_{c}=0$, $\tau_{\rm
dwell}/\tau_{\rm so}=0$ ($\beta=1$); $\theta_{0}/\theta_{c}=0$, $\tau_{\rm dwell}/\tau_{\rm so}=625$ ($\beta=4$); $\theta_{0}/\theta_{c}=0.4$,
$\tau_{\rm dwell}/\tau_{\rm so}=625$ ($\beta=2$). The values of the parameters
$c$ and $d$ for $\theta_0/\theta_c = 0$ are given by Eq.~\eqref{eq:cdparams}. For $\theta_0/\theta_c = 0.4$ we calculate $c = 0.427$, $d =
0.339$ from Ref.~\onlinecite{Mel97}.
The data is shown without any fit parameter in the lower panel and with $E_\text{gap}$,
$\Delta_\text{gap}$ as fit parameters in the upper panel. (A similar fitting procedure was used in Ref.~\onlinecite{Kor04}.) The fitted values of
$E_\text{gap}$ and $\Delta_\text{gap}$ do not vary much (by about 5\% and 10\%, respectively) from their nominal values [given by
Eq.~\eqref{eq:scalingpar}], but the agreement with the RMT predictions improves considerably if we allow for this variation.\cite{FOOTNOTE}

The characteristics of the gap distribution are clearly obtained in our
simulation. In zero magnetic field, the gap distribution becomes narrower as the
strength of spin-orbit coupling is increased ($\beta=1 \rightarrow 4$). The
RMT prediction for the standard deviation of the scaled distributions $P_\beta(x)$ is $\sigma_1 = 1.27$
and $\sigma_4 = 0.64$. The corresponding values obtained in our
numerical simulation {\em without any fitting} are  $\sigma_1 = 1.34$
and $\sigma_4 = 0.72$. If a weak magnetic field is present ($\beta=2$), 
the width of the distribution is predicted to be intermediate between the
cases with $\beta=1,4$. As seen, the dynamical model follows the
prediction. The theoretical value of the standard deviation is $\sigma_2 = 0.90$,
the numerical result (without fitting) is $\sigma_2 = 0.99$.

Using the same fit parameters as in Fig.~\ref{fig:P(x)} we plot the average density of
states close to the gap in
Fig.~\ref{fig:rhoscale(x)}. As seen, the numerical data follow closely
the analytical predictions, the deviations becoming significant only outside the
universal regime $|E-E_\text{gap}|\ll E_T$, i.e. $|x|\ll N^{2/3}$. In the absence of magnetic field  the spin-orbit
coupling induced oscillations are clearly obtained. 

The density of states in a strong magnetic field is given in
Fig.~\ref{fig:rhoCD}. The upper panel shows the data without spin-orbit coupling,\cite{Goo05} and the lower panel shows what happens if
spin-rotation symmetry is broken. The
 numerical data follows closely the analytical
prediction~\eqref{eq:rhoCD} of RMT. In particular the enhanced density of
 states  in the presence of spin-orbit coupling is clearly seen.
The first oscillations in the density of states are also captured in the dynamical model. The frequency doubling due to the reduced
degeneracy ($s=2 \rightarrow s=1$) is apparent.

\section{Conclusion}
In conclusion, we have introduced a quantum map for the dynamics of a chaotic quantum dot with spin-orbit coupling connected to a
superconductor. We have demonstrated three effects of spin-orbit coupling on the excitation spectrum of this Andreev billiard: The
narrowing of the distribution of the excitation gap and the appearance of oscillations in the density of states in the absence of a
magnetic field; and the peak in the density of states at the Fermi level in strong magnetic
field. Our numerical simulations confirm the predictions of random-matrix theory. The third effect is particularly interesting from an
experimental point of view. In view of the possibility to tune the strength of spin-orbit coupling in quantum dots,\cite{Zum02,Mil03} one can imagine
tuning the density of states at the Fermi level from zero to a value of twice the normal density of states.

\section*{ACKNOWLEDGMENTS}
Discussions and correspondence with J.~Cserti, M.~C.~Goorden, P.~Jacquod, and J.~Tworzyd{\l}o are gratefully acknowledged. This work was supported by the Dutch Science Foundation NWO/FOM. We acknowledge support by the European Community's Marie Curie
Research Training Network under contract MRTN-CT-2003-504574, Fundamentals of Nanoelectronics.


\begin{thebibliography}{22}
\bibitem[{FOO()}]{FOOTNOTE}
\bibinfo{note}{Ref.~\onlinecite{Jac03} reported good agreement for the
  $\beta=1$ gap distribution between RMT and the Andreev kicked rotator ---
  without any need to fit $E_\text{gap}$ and $\Delta_\text{gap}$. We now
  believe that this good agreement was fortuitous, and that some 5--10\%
  variation of these parameters is generally needed for a good agreement.}
\end{thebibliography}
\end{document}